\input{aipcheck}

\documentclass[
    ,final            
  ]
  {aipproc}

\layoutstyle{6x9}

\usepackage{epsf}

\newcommand{\be}{\begin{equation}} 
\newcommand{\ee}{\end{equation}}
\newcommand{\bea}{\begin{eqnarray}} 
\newcommand{\eea}{\end{eqnarray}}

\newcommand{\gton}{\mathrel{\lower.9ex \hbox{$\stackrel{\displaystyle 
>}{\sim}$}}} 
\newcommand{\lton}{\mathrel{\lower.9ex \hbox{$\stackrel{\displaystyle 
<}{\sim}$}}}

\newcommand{\vp}{{\vec p}}

\begin{document}

\title{Fate of the Mach cone in covariant transport theory}


\author{Denes Molnar}{
  address={Physics Department, Purdue University, 
           West Lafayette, IN 47907, USA},
  altaddress={RIKEN BNL Research Center, Brookhaven National Laboratory,
              Upton, NY 11973, USA}
}



\begin{abstract}
An intriguing potential signature of hydrodynamic behavior 
in relativistic $A+A$ reactions at 
Relativistic Heavy Ion Collider (RHIC)
energies is conical flow induced by fast supersonic particles traversing the 
hot and dense medium. 
Here I present first results on the evolution of
Mach shocks in $2\to 2$ covariant transport theory, 
in a static uniform medium.

\end{abstract}

\maketitle


\section{Introduction}

There has been a lot of recent theoretical interest in conical flow
in heavy-ion collisions\cite{StoeckerWP,ShuryakMach,NeufeldMach,BetzMach}.
Important open questions are 
how the collision dynamics (rapid expansion, phase transition, 
inhomogeneities) and dissipation (viscosity) affect this unique 
flow pattern. 

It would be natural to investigate these questions using
causal dissipative hydrodynamics, however, the 
3+1D codes required still have to be developed.
Here I employ a convenient alternative\cite{minvisc},
covariant transport theory, which
is fully causal and stable, can address full 3+1D, and 
has a well-defined hydrodynamic limit.

\section{Main Results}

I consider, as Refs. 
\cite{ZPC,Bin:Et,ZPCv2,nonequil,v2},
Lorentz-covariant on-shell Boltzmann transport theory
with elastic $2\to 2$ rates, but here couple the system to
an external current through the transport equation:
\be
p_1^\mu \partial_\mu f_1 = S(x, \vp_1) + C[f,f](x,\vp_1) 
+ C[f_{ext},f](x,\vp_1)
\label{BTE}
\ee
where 
\be
C[f,g] \equiv
\int\limits_3\!\!\!\!
\int\limits_4\!\!
\left(f_3 g_4 - f_1 g_2\right)
W_{12\to 34} \  \delta^4(p_1{+}p_2{-}p_3{-}p_4) \ .
\ee
The integrals are shorthands
for $\int_i \equiv \int d^3 p_i / (2E_i)$, 
the source term $S$ generates the initial conditions, while $f_{ext}$ 
represents the external current.
Because the interest is to study the theory 
{\em near its hydrodynamic limit},
I take $f$ and $f_{ext}$ to be the phasespace distribution of 
massless ``quasi-particles''. This ensures that the
equation of state $e=3p$ is close to that of the 
high-temperature plasma in the early stages at RHIC. 
The transition probability $W = s (s-4m^2) d\sigma/dt$
is adjusted to control the shear viscosity $\eta \approx 4T/(5 \sigma_{tr})$,
where $\sigma_{tr}$ is the transport cross section.

In this exploratory study I compute the flow pattern generated by
an external ``jet'' moving in the $+z$ direction 
in a static, uniform, thermal bath (a massless
gas of quarks and gluons - 3 flavors, 3 colors)
with temperature $T_0 = 0.385$ GeV 
and shear viscosity $\eta \approx 0.075 s$, 
where $s$ is the entropy density. 
The corresponding energy density and mean free path are
$e_0 \approx 44.7$ GeV/fm$^3$ and $\lambda = 0.125$~fm.
Though covariant transport 
treats interactions between jet and medium self-consistently,
I here explicitly turn off jet recoil, similar in spirit to linear response
studies. The jet is created at $t = 0$.

In the first 'perturbative' scenario 
the jet deposits energy and momentum through $2\to 2$ interactions
as encoded in (\ref{BTE}). This implies $dE/dL \approx dp_z/dL$ at high
jet energies $E \gg T$.
Typical Debye-screened $t-$channel processes are quite inefficient
at energy-momentum transfer to the medium, 
$dE/dL \sim (\mu_D^2/\lambda_{MFP}T) \ln (ET/\mu_D^2)$.
To maximize effects, I therefore take more optimistic 
{\em isotropic} scattering, for which
$dE/dL \sim E/(2\lambda_{MFP})$.
The jet is modeled through a moving sharp sphere profile
\be
f^{pert.}_{ext}(t,r,z,\vp) \ \propto\  
\frac{1}{R^3}\Theta(R^2 - (r^2 + (z-t)^2))
\, \delta^2(\vp_T)\, \delta(p_z - E)
\ee
of radius $R = 0.2$~fm, where $E$ is the jet energy,
and cylindrical $(r,z,\theta)$ coordinates are employed.

\begin{figure}
  \epsfysize=6.8cm
  \epsfbox{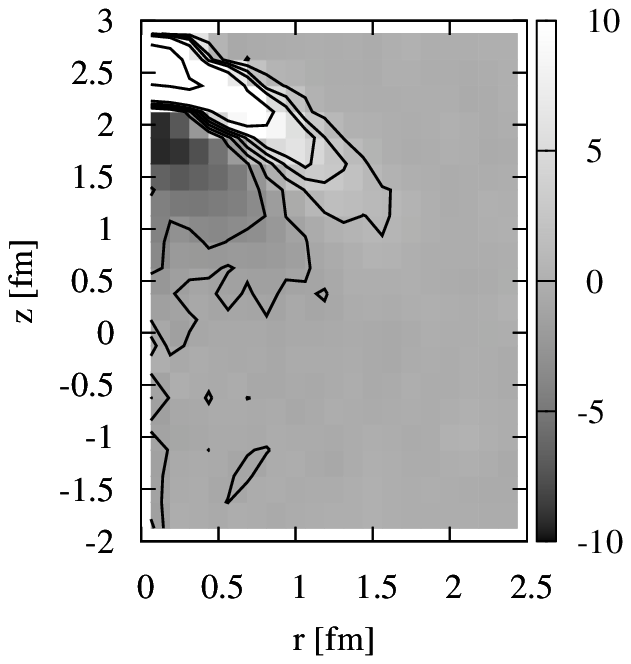}
  \epsfysize=6.8cm
  \epsfbox{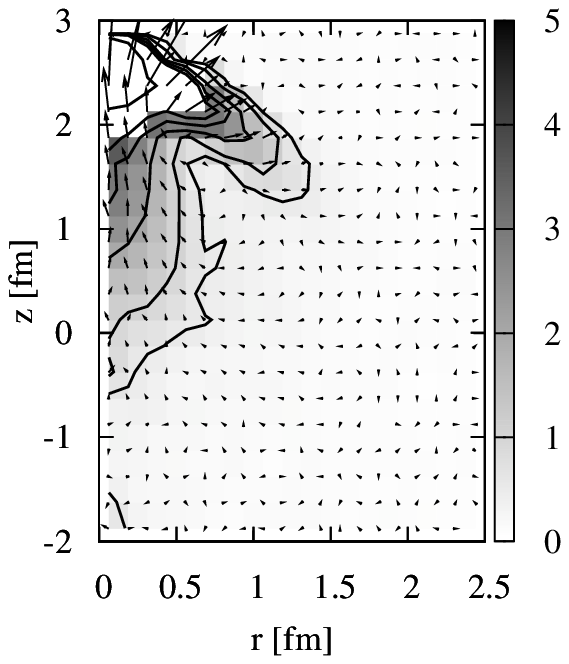}
\vspace*{-0.2cm}
  \caption{Distribution of the local energy density (left) and 
the magnitude of momentum density
perturbations (right), at $t = 2.5$~fm for the ``perturbative'' scenario.
Energy density contour lines are at 
$-3$, $-1$, $1$, $3$, $10$ and $40$ GeV/fm$^3$; 
while for the momentum density
at $0.5$, $1$, $2$, $3$, $4$, and $20$~GeV/fm$^3$.
Arrows indicate the local momentum density vectors.}
\label{Fig:1}
\end{figure}

Figure~\ref{Fig:1} shows the jet-induced {\em change} in energy density and 
momentum density in the medium, at $t = 2.5$~fm, for $E=8.25$~GeV
and $dE/dL \approx dp_z/dL \approx 20$~GeV/fm$^3$. The largest
effect is, of course,
right at the position of the jet but there clearly is a bow-shaped front
trailing the jet. In the wake of the jet
($z\lton 2$~fm, $r \lton 0.3$~fm), 
momentum flows along the jet direction (characteristic
``diffusion wake'').
These results agree qualitatively with ideal hydrodynamic 
calculations\cite{BetzMach}.
This is quite remarkable because in {\em this} calculation
energy-momentum deposition is not thermal, and there is also a finite,
albeit small viscosity.

As an alternative scenario, consider a source that deposits 
thermalized energy only but no momentum ($dp_z/dL = 0$). This 
can be incorporated
through additional thermal particle production, i.e.,
adding to the source term $S$ in (\ref{BTE}) a contribution
\be
\Delta S(t,r,z,\vp) \ \propto \  v\, \frac{dE}{dL}\, 
e^{-[r^2+(z-vt)^2]/(2\sigma^2)}
\, \, e^{-p/T_0}
\ee
while putting $f_{ext} = 0$.
Here a Gaussian spatial 
profile was chosen with a width $\sigma = 0.3$~fm, and $v$
is the source velocity. 

\begin{figure}
  \epsfysize=6.8cm
  \epsfbox{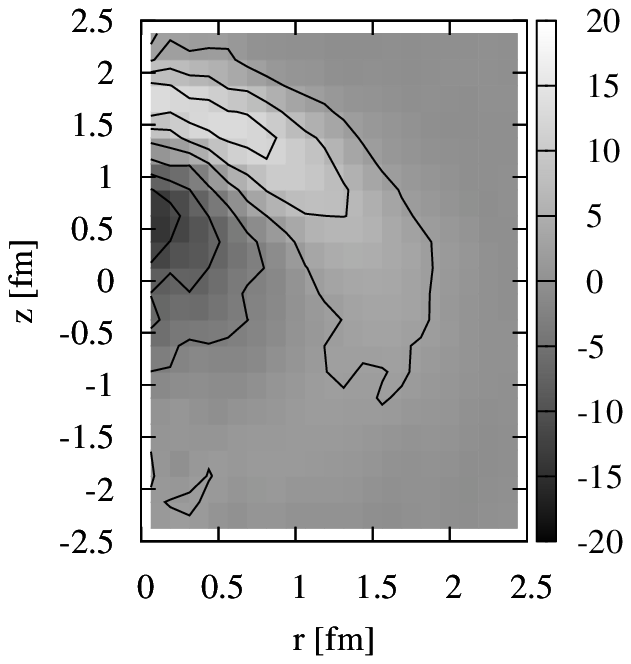}
  \epsfysize=6.8cm
  \epsfbox{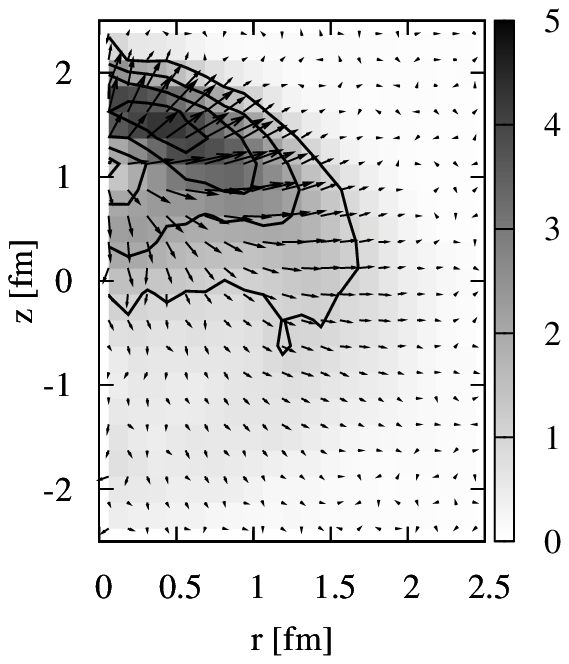}
\vspace*{-0.2cm}
  \caption{Same as Fig.~1 but for the ``pure energy'' scenario,
with contour lines
at $-12.5$, $-7.5$, $-2.5$, $2.5$, $7.5$, and $12.5$~GeV/fm$^3$
for energy density and $1$, $2$, $3$, and $4$~GeV/fm$^3$
for momentum density.}
\label{Fig:2}
\end{figure}

Figure~\ref{Fig:2} shows the {\em change} in energy density and 
momentum density in the medium, at $t = 2.5$~fm, 
in the ``pure energy'' scenario for $dE/dL \approx 75$~GeV/fm$^3$
and $v = 0.9c$. Because the source is turned off for
$t > 2$ fm,
the bow-shock is now more pronounced. 
In contrast to the ``perturbative'' scenario,
in the wake region we see momentum flow {\em away} from the jet.
In the regions with the highest momentum density, the flow angle is
consistent with $\approx 50$ degrees expected from the Mach formula
$\cos \theta = c_s/v$ (here $c_s^2 = 1/3$).
These features agree 
qualitatively with ideal hydrodynamic calculations\cite{BetzMach}.

\begin{figure}
\leavevmode
  \epsfysize=6.8cm
  \epsfbox{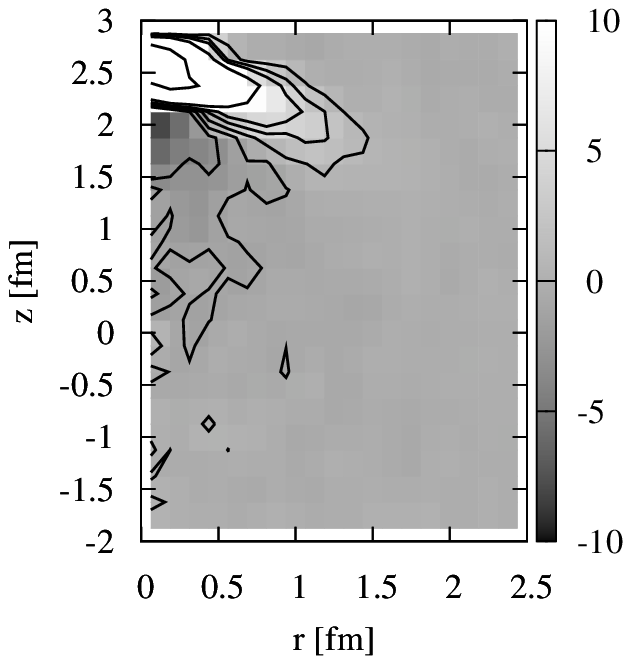}
  \epsfysize=6.8cm
  \epsfbox{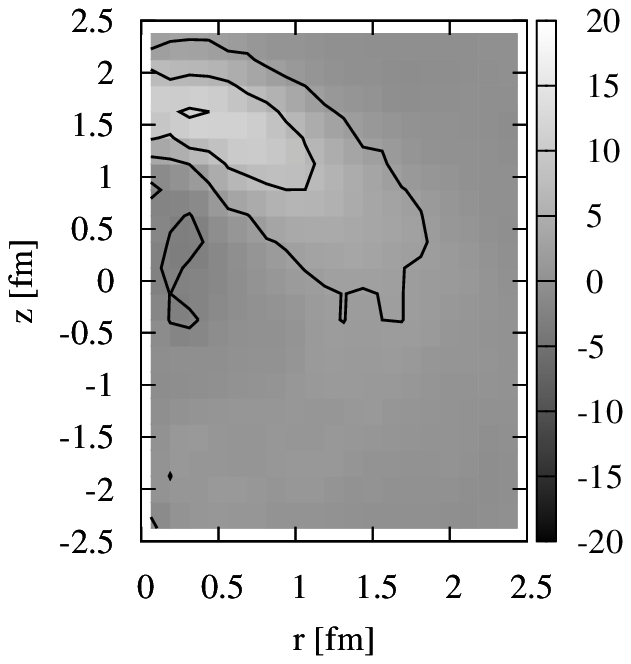}
  \caption{Distribution of the local energy density perturbations
for the ``perturbative'' (left) and ``pure energy'' scenarios (right),
same as Figs.~\ref{Fig:1}(left) and Figs.~\ref{Fig:2}(left),
but with four times larger mean free path.}
\label{Fig:3}
\end{figure}

For bow shocks to manifest, a low viscosity is crucial.
As seen in Fig.~\ref{Fig:3}, the disturbances get largely ``washed out''
if the shear viscosity is quadrupled to $\eta \approx 0.3 s$.

For the modest number of test particles and events
in this exploratory study,
there were no clear signals in
the azimuthal distribution in the $x-z$ plane, $dN/dy\, d\phi$, 
and the angular distribution relative to the jet axis $dN/d\cos\theta$
(even when weighted by powers of momentum).
Ideal hydrodynamic calculations also find very small effects,
confined to thermal tails $p/T \sim 20$ \cite{BetzMach}.
It remains to be seen whether the dynamics and self-consistent
jet-medium coupling can generate appreciable signals
in heavy-ion collisions.

\section{Conclusions}

In this work I investigate
Mach shocks in $2\to 2$ covariant transport theory, 
in a static uniform medium. 
If the shear viscosity to entropy density ratio is very low,
$\eta/s \approx 0.075$,
the results are in qualitative agreement with
ideal hydrodynamic calculations\cite{BetzMach}.
This demonstrates the feasibility of utilizing covariant transport in 
future 3+1D conical flow studies to incorporate viscosity and
a self-consistent coupling between jet and medium.


\begin{theacknowledgments}
I thank RIKEN, 
Brookhaven National Laboratory and
the US Department of Energy [DE-AC02-98CH10886] for providing facilities
essential for the completion of this work.
\end{theacknowledgments}


\end{document}